\pdfoutput=1

\documentclass[11pt,twoside,a4paper,cmspaper,final,collab]{cms-tdr}
\def\svnVersion{6c7d37c-D}\def\svnDate{2020/04/21}\def\cmsCernNoTag{CERN-EP-2020-004}\def\cmsCernDate{\today}\def\cmsMessage{"Published in Physics Letters B as \href{http://dx.doi.org/10.1016/j.physletb.2020.135425}{\doi{10.1016/j.physletb.2020.135425}.}"}

\begin{document}\cmsNoteHeader{HIG-19-004}

\hyphenation{had-ron-i-za-tion}
\hyphenation{cal-or-i-me-ter}
\hyphenation{de-vices}
\RCS$HeadURL: svn+ssh://svn.cern.ch/reps/tdr2/notes/HIG-16-040/trunk/HIG-16-040.tex $
\RCS$Id: HIG-16-040.tex 418438 2017-07-28 15:48:53Z sethzenz $
\newlength\cmsFigWidth
\ifthenelse{\boolean{cms@external}}{\setlength\cmsFigWidth{0.98\columnwidth}}{\setlength\cmsFigWidth{0.65\textwidth}}
\ifthenelse{\boolean{cms@external}}{\providecommand{\cmsLeft}{top\xspace}}{\providecommand{\cmsLeft}{left\xspace}}
\ifthenelse{\boolean{cms@external}}{\providecommand{\cmsRight}{bottom\xspace}}{\providecommand{\cmsRight}{right\xspace}}
\newlength\cmsTabSkip\setlength{\cmsTabSkip}{1ex}
\cmsNoteHeader{HIG-19-004} 

\title{A measurement of the Higgs boson mass in the diphoton decay channel}

\date{\today}
\newcommand*{\mgg}{\ensuremath{m_{\gamma\gamma}}\xspace}
\newcommand{\Ztoee}{\ensuremath{\PZ\to\Pe\Pe}\xspace}
\newcommand{\Ztomumu}{\ensuremath{\PZ\to\PGmp\PGmm}\xspace}
\newcommand*{\Zee}{\ensuremath{\PZ \to \Pe\Pe\xspace} }
\newcommand{\hgg}{\ensuremath{ \PH \to \gamma \gamma}\xspace }
\newcommand{\hzz}{\ensuremath{ \PH \to \PZ\PZ \to 4\ell} }
\newcommand*{\gjet}{$\gamma$ + jet }
\newcommand{\ggH}{\ensuremath{\Pg\Pg\PH}}
\newcommand{\ttH}{\ensuremath{\ttbar\PH}}
\newcommand{\VH}{\ensuremath{\PV\PH}}
\newcommand{\mH}{\ensuremath{m_{\PH}}\xspace}
\newcommand{\mee}{\ensuremath{m_{\Pe\Pe}}\xspace}
\newcommand{\Rn}{\RNINE\xspace}

\abstract{
A measurement of the mass of the Higgs boson in the diphoton decay channel is
presented.
This analysis is based on 35.9\fbinv of proton-proton collision data
collected during the 
2016 LHC running period, with the CMS detector
at a center-of-mass energy of 13\TeV. A refined detector calibration
and new analysis techniques have been used to improve the precision of this
measurement.
The Higgs boson mass is measured to be $\mH = 125.78 \pm 0.26\GeV$.
This is combined with a measurement of \mH already performed
in the \hzz decay channel using the same data set, giving  
 $\mH = 125.46 \pm 0.16\GeV$. This 
result, when further combined with an earlier measurement of \mH
using data collected in 2011 and 2012 with the CMS detector, gives a value for the Higgs boson mass of $\mH = 125.38 \pm 0.14\GeV$. 
This is currently the most precise measurement of the mass of the
Higgs boson.
}

\hypersetup{%
pdfauthor={CMS Collaboration},%
pdftitle={A measurement of the Higgs boson mass in the diphoton decay channel},%
pdfsubject={CMS},%
pdfkeywords={CMS, Higgs, diphoton}}

\maketitle 

\section{Introduction}
\label{sec:intro}
The independent observations of the Higgs boson
by the ATLAS and CMS
Collaborations~\cite{Aad:2012tfa,Chatrchyan:2012ufa,CMSLONG}
in proton-proton collisions at the CERN LHC was a key milestone in the
understanding of
the mechanism of electroweak symmetry breaking.
More recently, with the increased amount of data resulting from the higher
energy and the higher luminosity accumulated
at the LHC between 2015 and 2018 (Run 2), the focus has shifted from
observation to precision measurements of its properties.
The couplings of the Higgs boson to other elementary particles
can be predicted by the standard model of particle physics once its mass is
known.
This motivates precise measurements of the mass of the Higgs boson (\mH) in
all available decay channels.

Although the \hgg decay channel has a small ($\approx$0.23\%) branching
fraction, it provides a clean final state topology in which the diphoton
invariant mass can be reconstructed with high precision.
The measurement of \mH in this decay channel can be combined with measurements
in other decay channels to achieve an even higher precision. In this way the ATLAS and CMS Collaborations measured
\mH to be $125.09\pm 0.24\GeV$~\cite{Aad:2015zhl} 
 with the data collected in 2011 and 2012 (Run 1).

In this Letter, we present a new measurement of \mH in the \hgg decay channel
with the
 data collected at $\sqrt{s} = 13\TeV$ in 2016 corresponding to an integrated
luminosity of 35.9\fbinv. The CMS Collaboration has
previously reported a measurement of \mH  with the same data set in the
\hzz decay channel where \mH was measured to be $125.26\pm 0.21\GeV$~\cite{Sirunyan:2017exp}. 
The ATLAS collaboration have also published a 
measurement of \mH of $124.97\pm 0.24\GeV$~\cite{Aaboud:2018wps}, using the
combined 2016 and Run 1 data sets.
Our measurements of
\mH with the 2016 data set, in the \hgg and \hzz decay channels, have been combined with our measurement of \mH
 with the Run 1 data set.
The combined result and the procedure followed for this combination are also
described in this Letter.

\section{The CMS detector}
\label{sec:detector}

The central feature of the CMS detector is a superconducting solenoid of
6\unit{m} internal diameter with a uniform magnetic field of 3.8\unit{T}.
Inside
the magnet volume are silicon pixel and strip trackers, a lead tungstate
crystal electromagnetic
calorimeter (ECAL), and a brass and scintillator hadron calorimeter, each
composed of a barrel
 and two endcap sections. Gas-ionization chamber based muon detectors are
embedded in the steel flux-return yoke outside the solenoid.
The ECAL is a hermetic homogeneous calorimeter made of
61\,200 lead tungstate (PbWO$_4$) crystals mounted in the central barrel part,
closed by 7324 crystals
in each of the two endcaps. In the region $1.65 < \abs{\eta} < 2.6$ a
three-radiation-length-thick preshower
detector with two orthogonal layers of silicon strips
is placed in front of the endcap crystals. Avalanche photodiodes are used as
photodetectors in the barrel
and vacuum phototriodes in the endcaps. The barrel part of the ECAL (EB)
covers the pseudorapidity
range  $\abs{\eta} < 1.479$, while the endcap calorimeters cover the range $1.479
< \abs{\eta} < 3.0$.
 A calorimeter with longitudinal quartz fibres complements the coverage
provided
by the barrel and endcap detectors.
The first level of the CMS trigger system~\cite{Trigger} uses information
from the calorimeters and muon detectors to select the most interesting events
in a fixed time interval
of less than 4\mus. The high-level trigger processor farm further decreases
the event rate from around 100\unit{kHz} to
around 1\unit{kHz} before data storage.
A more detailed description of the CMS detector, together with a definition of
the coordinate system and the relevant kinematic variables, can be found in
Ref.~\cite{Chatrchyan:2008zzk}.

\section{Analysis strategy}
\label{sec:strat}

The general strategy followed in this analysis is the same as that adopted in
an earlier analysis by the CMS Collaboration of the Higgs boson properties in
the diphoton channel~\cite{HIG-16-040}.
Since that publication, refinements were made to increase the  precision of
the measurement of
\mH through a better understanding of the systematic uncertainties of the
measurement, and
a more accurate detector calibration was performed.
We have also improved the method, first introduced in Ref.~\cite{Run1HGG}, to
measure and correct for nonlinear discrepancies in the energy scale  with
transverse momentum (\PT), of electrons from \PZ boson decay, between data and
simulation by increasing the granularity of the correction.
In addition, we have developed a method to evaluate the systematic uncertainty
of the photon energy scale due to radiation damage of the ECAL crystals, and a
simplified event categorisation, described in Section 6, is followed in the analysis.

With the new calibration, the detector response is more stable with time,
leading to a reduction of the uncertainties in the corrections to the photon
energy due to the material upstream of the ECAL and of the uncertainties
associated with variables which describe the electromagnetic shower.

\section{Data and simulation}
\label{sec:samples}

The events used in this analysis were collected in 2016 with an 
integrated luminosity of 35.9\fbinv. They were selected with a diphoton trigger that had
asymmetric \pt thresholds of 30 and 18\GeV.
Full details of the trigger selection and the measurement of the trigger
efficiency can be found in Ref.~\cite{HIG-16-040}.
To model the signal and background processes, events are generated with  Monte Carlo techniques. The detailed response of the CMS detector is simulated
using the \GEANTfour
package~\cite{Agostinelli:2002hh}.

Signal events are simulated with the \MGvATNLO v2.2.2 matrix-element
generator~\cite{AMCAT} at next-to-leading order and interfaced with
\PYTHIA8.205~\cite{Pythia8} for parton showering and hadronization. The
\PYTHIA underlying event tune CUETP8M1~\cite{CUETP8} was used.
The irreducible prompt diphoton background and the reducible backgrounds of
\gjet and multijet events, where the jets are misidentified as
isolated photons, are the dominant backgrounds to the \hgg decay process. The
diphoton background is modelled with the \SHERPA v.2.2.1~\cite{SHERPA}
generator, which includes the Born processes with up to 3 additional jets at
leading order (LO) accuracy, as well as the LO box processes. The
$\gamma$+jets and multijet backgrounds are modelled with \PYTHIA at LO. 
These samples are used for the training of the multivariate discriminants used
in this analysis, as well as for the optimisation of the event categorisation.
The Drell--Yan samples used to derive the electron and photon energy scale
corrections and their systematic uncertainties, are simulated with
\MADGRAPH~\cite{Mad5} and \MGvATNLO generators and merged together in order to
improve the statistical precision of the scale corrections. Before merging these 
samples, the compatibility of the \mee lineshapes between the two generators 
in the categories used to derive the electron and photon energy scale corrections was confirmed.

The simulation includes multiple proton-proton interactions taking place
within a bunch crossing, known as `pileup'. 
Pileup can occur not only in the
same bunch crossing (in-time pileup), but also in the crossing of previous and
subsequent bunches (out-of-time pileup), both of which are accounted for by the simulation.
The simulated events are scaled to reproduce the distribution of 
the number of pileup interactions in data.

\begin{figure}[t]
\centering
\includegraphics[width=\cmsFigWidth]{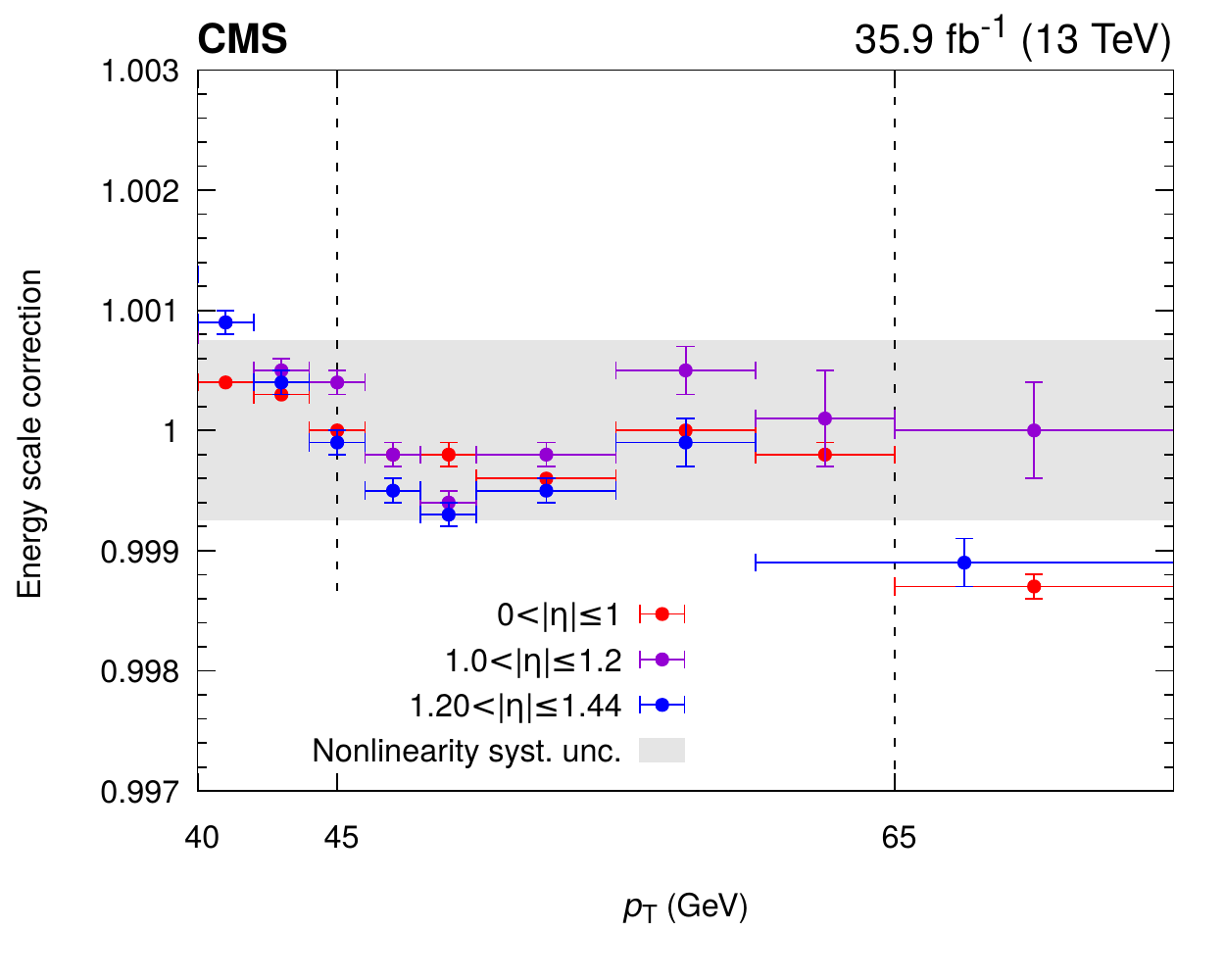}
\caption{Energy scale corrections as a function of the \PT of the photon. The
horizontal bars in the plot represent the variable bin width. The
systematic uncertainty associated with this correction is approximately the
maximum deviation observed in the \PT range between 45 and 65\GeV for
electrons in the EB region.}
\label{fig:scale_vs_et}
\end{figure}

\begin{figure*}
    \centering
    \includegraphics[width=0.45\textwidth]{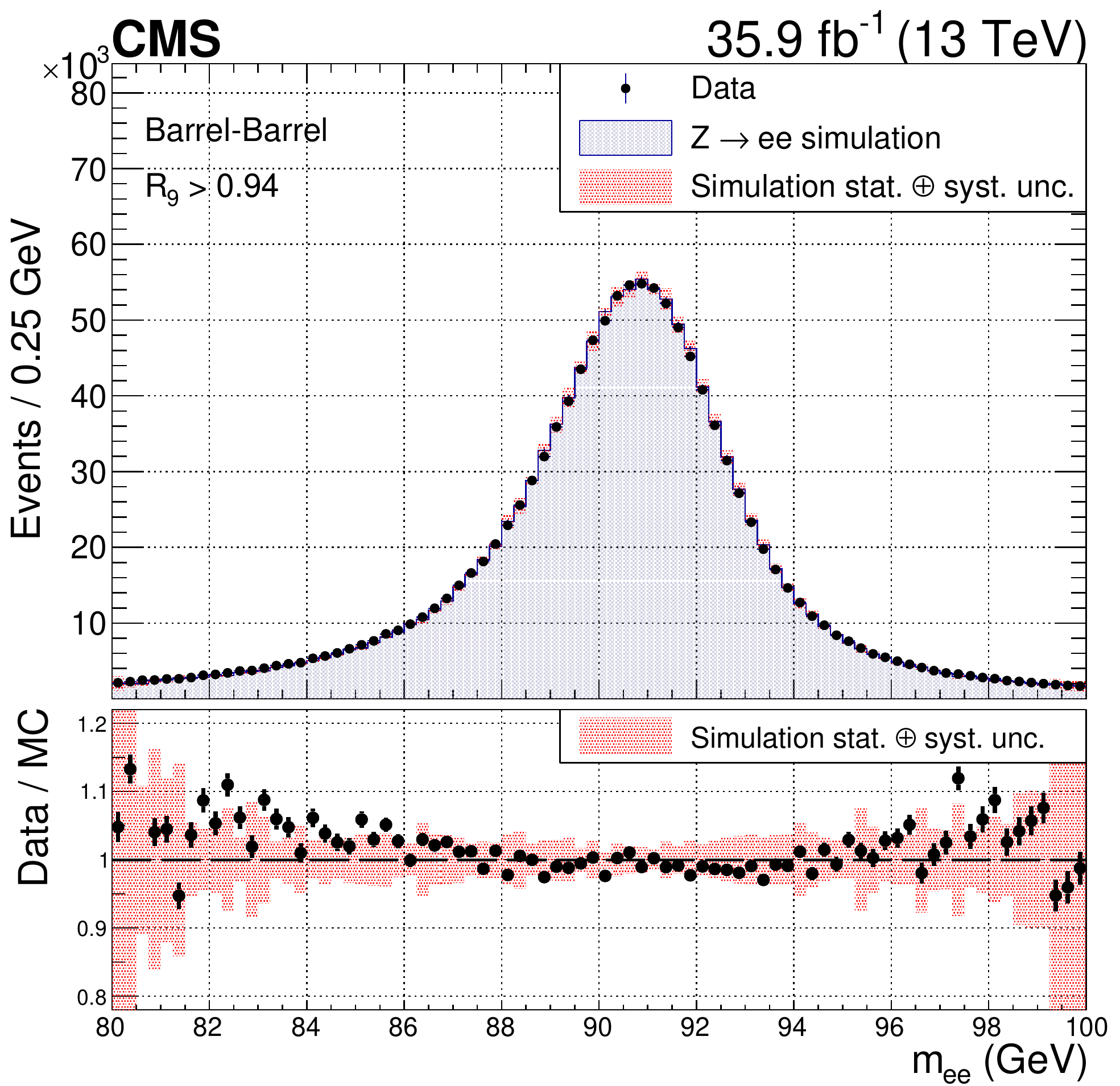}
    \includegraphics[width=0.45\textwidth]{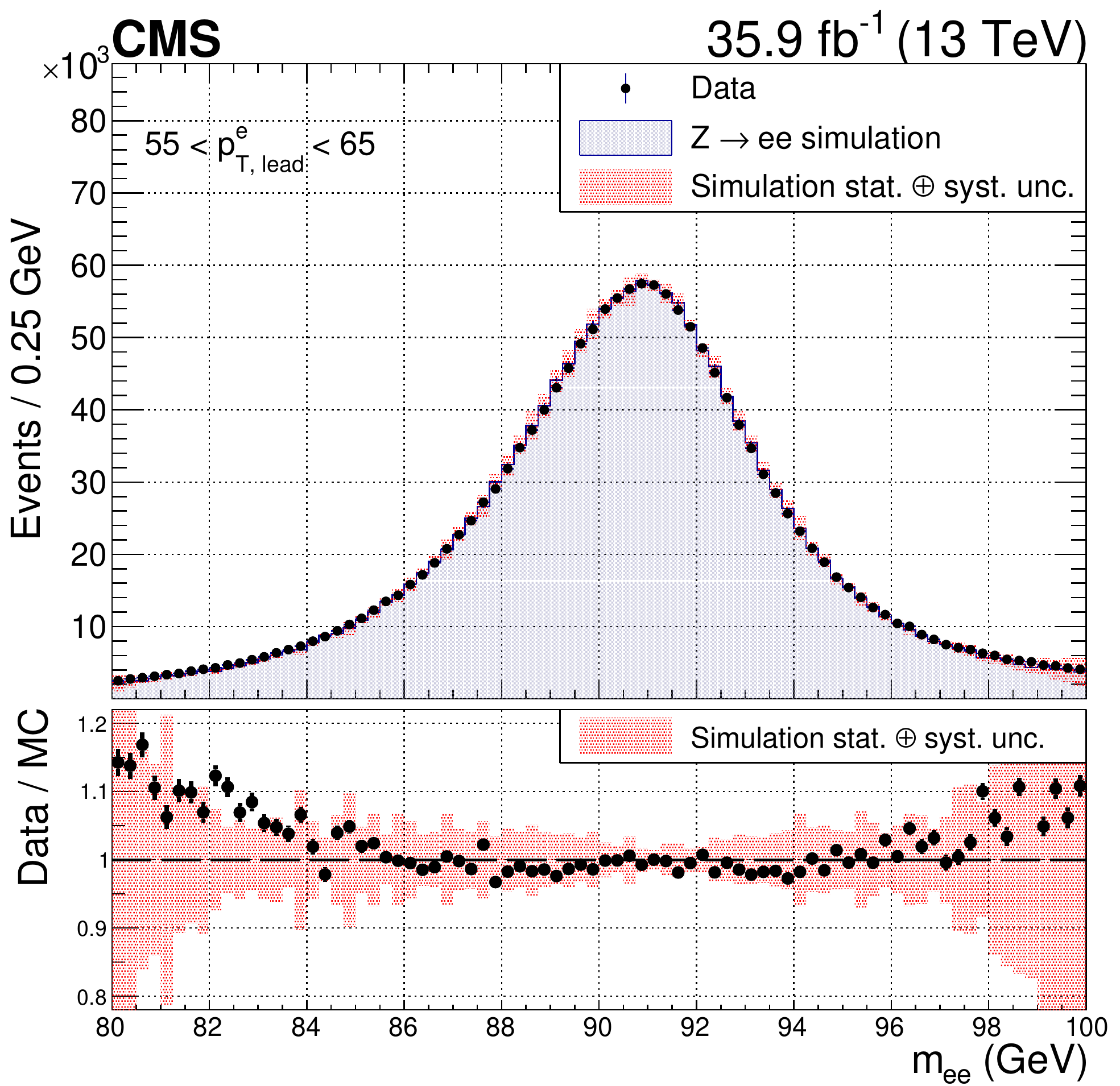}
    \caption{Comparison of the  distributions of the invariant mass of the
dielectrons in data and simulation in \Ztoee events after
    application of energy corrections in two representative
    categories.
    Left: Both electrons are in the EB and satisfy $\Rn > 0.94$.
    Right: the leading electron has a transverse momentum between 55 and
65\GeV, without a requirement on the second electron. The systematic
uncertainty in the error band in the plots include only the uncertainties on
the derived energy scale corrections. }
    \label{fig:inv_mass_agreement}
\end{figure*}

\section{Photon reconstruction and identification}
\label{sec:phoRecoAndID}

Photon candidates are reconstructed as energy deposits in a collection of
crystals in the ECAL.
A cluster is formed by first identifying a `seed'  crystal with an energy
above a given threshold,
then the cluster is built by finding the crystals that share an edge with the
seed crystal and have an energy above another, lower threshold. This second
threshold is set to be approximately 80\MeV in the barrel and ranging
from 80 to 300\MeV in the endcaps, depending on $\abs{\eta}$. These
clusters, once formed, are combined to form a `supercluster', aiming to fully
contain the shower of the photon. This procedure accounts for variations in
geometry as a function of $\abs{\eta}$, and optimises the robustness of the
energy resolution against pileup.

\subsection{Photon energy calibration}

A critical component of the measurement of \mH is the energy calibration of
the response of the ECAL to photons.
    The energy of a photon is calculated by summing the calibrated and
corrected energy~\cite{Supercluster} of all crystals in the associated
supercluster, and the energy deposited in the preshower in the region 
$1.65 <\abs{\eta} < 2.6$ covered by this detector.
For each supercluster, a shower shape variable \Rn is defined, which is used
to select photons undergoing a conversion in the material between the
interaction point and the front face of the ECAL. The variable \Rn is defined
for a candidate electromagnetic cluster as the ratio of the sum of energy
deposited in a $3{\times}3$ crystal array, centred on the crystal with the
highest energy, to the sum of the energy in the supercluster. The energy
deposition of photons that convert before reaching the calorimeter tends to
have wider transverse profiles and thus lower values of \Rn than those of
unconverted photons.
To further optimise the energy resolution, the energy is corrected for the
lack of complete containment of the electromagnetic showers in the clustered
crystals, the energy lost by photons that convert upstream of the calorimeter,
and the effects of pileup. These corrections are derived using a multivariate
regression technique, trained on simulated events, which simultaneously
estimates the energy of the photon and its median uncertainty. The inputs to
this regression are shower shape variables, the preshower information, and
observables sensitive to pileup~\cite{Regression}.

    After applying these corrections to the photon energy, some residual
differences remain between the data and simulation in both the photon energy
scale and the resolution. A multistep procedure is used to correct these
differences, using \Ztoee decays in which the electron showers are
reconstructed as photons,
 so that the simulation accurately reproduces the data.
In the first step of this process, any residual long-term drifts in the energy
scale in data are corrected for, in approximately 18-hour intervals
corresponding to one LHC fill.
In the second step,
corrections to both the
energy resolution in the simulation,  and the scale correction needed for the
data are derived simultaneously in bins of $\abs{\eta}$ and \Rn for electrons.
The energy resolution obtained in simulation is matched to the data by adding
a Gaussian smearing term, determined by adjusting the agreement in the \Ztoee
invariant mass distributions. In the third and final step the energy scale
corrections are derived in bins of $\abs{\eta}$ and \PT to account for any
nonlinear response of the crystals with energy.
The corrections obtained from this step are shown in
Fig.~\ref{fig:scale_vs_et} for electrons as a function of \PT in the three
bins of $\abs{\eta}$ in EB.
This additional step in the scale correction improves the precision of the
measurement of \mH, since the energy spectrum of the electrons from \PZ boson
decay ($\langle\PT\rangle\approx 45\GeV$) used to derive the scale
corrections, is different from the energy spectrum of photons from Higgs boson
decay ($\langle\PT\rangle\approx 60\GeV$).

We note that in the second step the number of bins in \Rn for the scale
corrections has been
increased by a factor of five over the previous analysis~\cite{HIG-16-040},
resulting in an improvement in the
precision with which the energy scale is determined.  Also, in order to
provide a consistency test of the derivation procedure, the
correction factors that are obtained in the second and third steps are applied a second time
to the data and a new set of
factors is extracted in the same electron categories. Any deviation from unity is an indication of the
nonclosure of the derivation procedure and is applied
as a systematic uncertainty on scale corrections.

The agreement between data and simulation in the dielectron invariant mass,
after applying these energy scale corrections and the additional smearings, is
shown in Fig.~\ref{fig:inv_mass_agreement} for dielectron events in the EB
with \Rn greater than 0.94, and for dielectron events with a leading
transverse momentum between 55 and 65\GeV, without a requirement on the second
electron. The former demonstrates the performance of the energy corrections on
photons with the highest event count, optimal resolution, and the highest
sensitivity to the Higgs boson mass. The latter demonstrates that the energy
corrections are effective in a kinematic region where the \PT of the electron
has been chosen to be the typical \PT of a photon from a Higgs boson decay. In
both cases data and simulation are in good agreement in the core of the
distributions.

\subsection{Photon preselection and identification}

The photons  considered in the subsequent steps of this analysis are required
to satisfy certain preselection criteria that are similar to, but more
stringent than, those imposed by the trigger requirements. A detailed
description of these preselection criteria, as well as the methods employed to
evaluate their efficiencies, can be found in Ref.~\cite{HIG-16-040}.
A dedicated boosted decision tree (BDT) is used to classify prompt photons
from other photon candidates that
arise out of misidentified jet fragments, but which satisfy the preselection
criteria. The full details of the input features
 of this photon identification BDT is also described in
Ref.~\cite{HIG-16-040}. The score of this BDT is used later
 in the event categorization, discussed in the next section.

\subsection{Vertex selection}

The identification of the diphoton vertex position along the beam axis has a
direct impact on the diphoton mass resolution, since if the vertex position is
known to better than about 1\cm, then the invariant mass resolution is
dominated by the photon energy resolution.
The distribution of the position of the interaction vertices along the beam
axis has an RMS spread of about 3.4\cm, and, in typical pileup
conditions in 2016, there were on average around 23 interactions in each bunch
crossing.
The choice of the diphoton vertex is made following the same procedure in
Ref.~\cite{HIG-16-040}: a BDT, whose inputs are observables related to tracks
recoiling against the diphoton system, is used to identify the most likely
vertex. A second BDT is used to determine the probability of correctly
choosing that vertex. The score of the second BDT is used later in the event
categorisation, discussed below.
The algorithm is validated using
\Ztomumu events with the muon tracks removed so as to mimic diphoton pair
production.
The efficiency of assigning the event to a vertex within 1\cm of the true
vertex in the simulated \hgg events is found to be approximately 81\%.

\section{Event classification}
\label{sec:categories}

The event selection procedure is similar to that in Ref.~\cite{HIG-16-040}.
The \PT of the two leading photons ($\PT^{\gamma1}$, $\PT^{\gamma2}$) are
required to satisfy $\PT^{\gamma1} > \mgg/3$ and $\PT^{\gamma2} > \mgg/4$, where \mgg is the diphoton mass, and
the photon \PT requirement is applied after the vertex assignment.
Additionally \mgg is required to be between 100 and 180\GeV. The use of \PT
thresholds scaled with the diphoton invariant mass is to prevent a distortion
of the lower end of the invariant mass spectrum.
  The superclusters of both photons are required to have $\abs{\eta} < 2.5$
and to be outside of the barrel-endcap transition region, $1.44 < \abs{\eta}\leq 1.57$.

To improve the sensitivity of the analysis, events are classified according to
their  production mechanism, mass resolution, and their predicted
signal-to-background ratio.
A dedicated classifier, referred to as the diphoton BDT, is used to
discriminate between signal and background events. This BDT assigns a high
score to events with photons exhibiting signal-like kinematics, a good mass
resolution, and a high score from the photon identification BDT. The per-event
probability estimate of assigning the correct primary vertex to the diphoton
system is used as one of the input features of this diphoton BDT. The other
input features are described in Ref.~\cite{HIG-16-040}.

Nearly 95\% of Higgs boson events come from two production modes. These are
gluon-gluon fusion ($\cPg\cPg\PH$) and vector boson fusion (VBF), where there
are two jets in the final state separated by a large rapidity gap. A
multivariate discriminant is trained to discriminate VBF events from \ggH +
jets events, using the kinematics of the characteristic VBF dijet system as
inputs. This discriminant
is then given as an input to an additional multivariate classifier (VBF
combined BDT) along with the
score from the diphoton BDT, and the ratio $\PT^{\gamma\gamma}/\mgg$. The VBF
events are subdivided into three categories
based on the VBF combined BDT score. The remaining events are mostly
$\cPg\cPg\PH$ events and are designated as `untagged'.
These events are further subdivided into four categories based on their
diphoton BDT score. 

Adding other possible analysis categories, where for example, the Higgs boson
is produced in association with a vector boson, or with a pair of top quarks,
adds only a small increment to the precision of the mass measurement at the
cost of a
significant increase in the analysis complexity. Thus, unlike in the earlier analysis~\cite{HIG-16-040}, these production modes
are not considered as separate categories in this analysis.

\section{Signal and background models}
\label{sec:statAnalysis}

In order to extract \mH, signal and background models are constructed to fit
the diphoton mass distributions observed in the data.
The signal models are derived using simulated Higgs boson events, while the
background models used in the fits of the \mgg spectra are derived directly
from data.

\subsection{Signal model}
\label{sec:results_signal}

The resolution of \mH in the diphoton decay channel depends on the production
mechanism and the analysis category. Hence
the signal shapes used to model the diphoton invariant mass distributions are
derived for every analysis category and with a nominal value for \mH, using
simulated events from the different production modes.
The simulation accounts for the trigger, reconstruction, and identification
efficiencies, which are measured with data-driven techniques.
A weight is applied to the simulated events so that the distribution of the
number of interactions per bunch crossing and the location of the primary
vertex are matched to the distributions observed in data. A detailed
description of each of these steps can be found in Ref.~\cite{HIG-16-040}.

Since the distribution of \mgg depends on the correct assignment of the vertex
associated with the diphoton candidate,
signal models were constructed with correct and wrong vertex assignment
scenarios separately. For each process, analysis category, and vertex
scenario, the $\mgg$ distributions were fit with a sum of, at most, four
Gaussian functions.

For each process, analysis category, and vertex scenario, a simultaneous fit
of the signal samples at mass values ranging from 120 to 130\GeV is performed
to obtain the variations of the parameters of the Gaussian functions, described
by polynomials in \mH, used in the signal model fit.

\begin{figure*}
  \centering
  \includegraphics[width=0.43\textwidth]{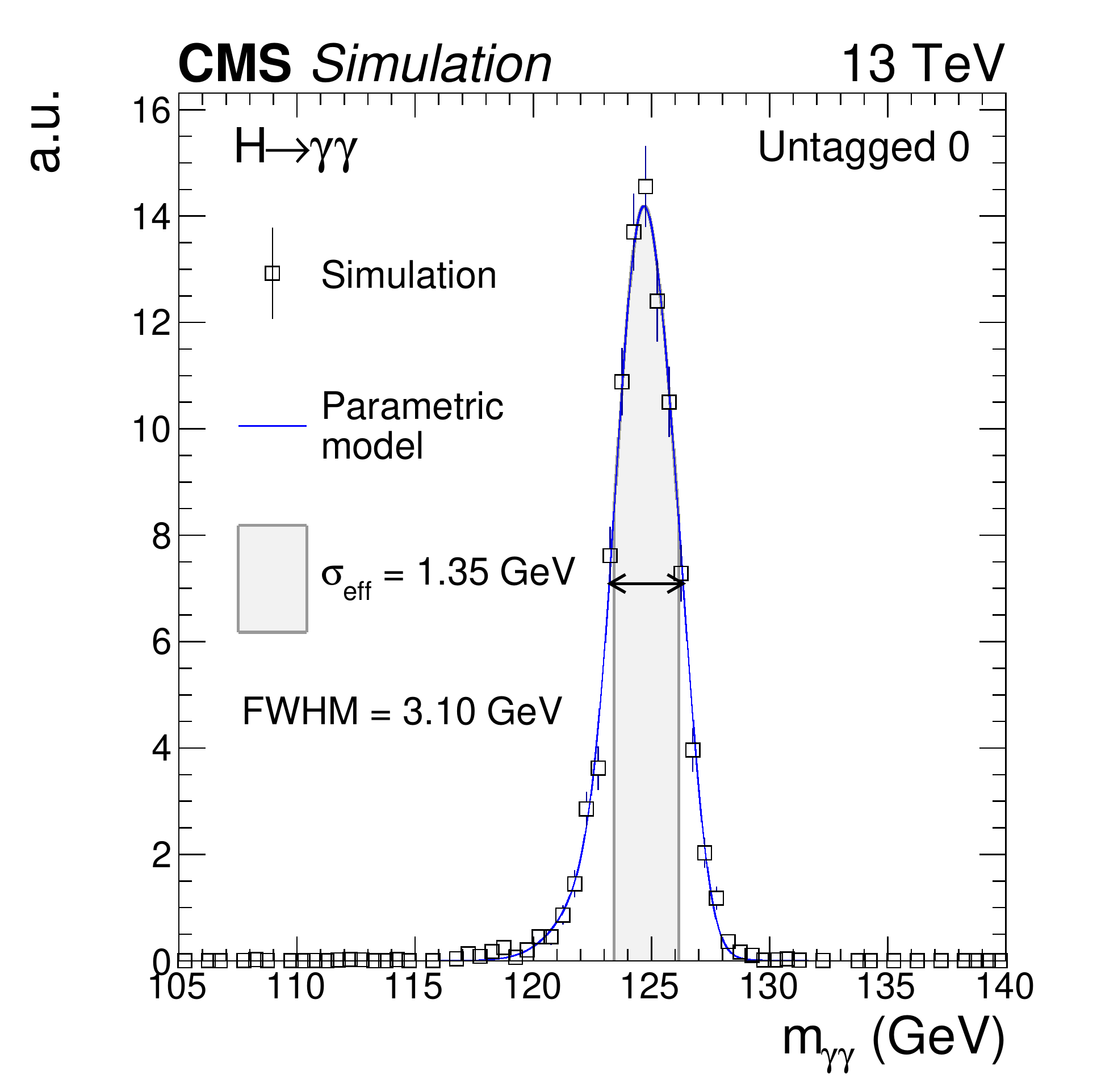}
  \includegraphics[width=0.429\textwidth]{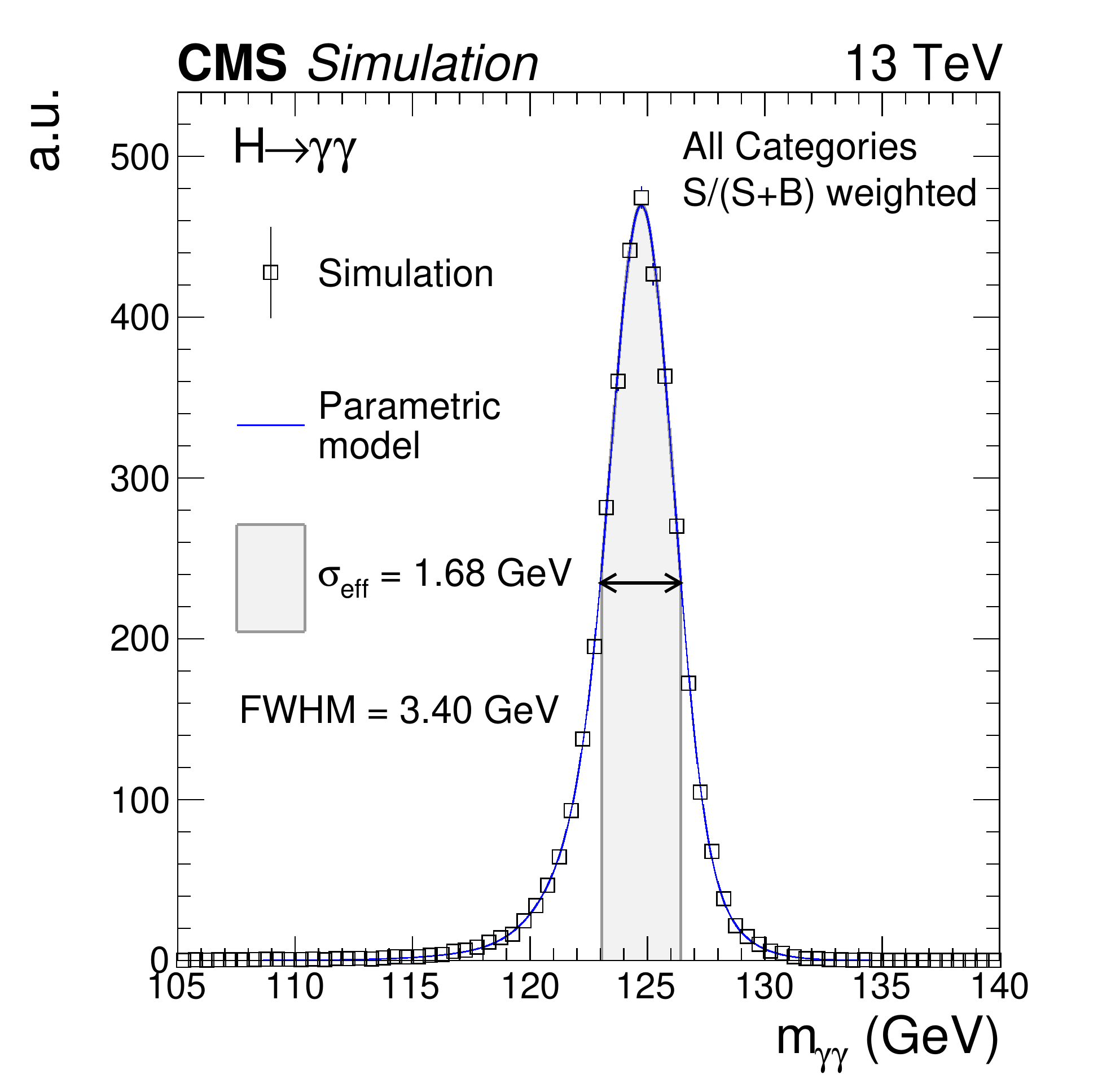}
  \caption{The signal shape models for the highest resolution analysis
category (left), and the sum of all categories combined together after scaling
each of them by the corresponding S/(S+B) ratio (right) for a simulated \hgg
signal sample with $\mH = 125\GeV$. The open squares represent weighted
simulated events and the blue line represents the
corresponding model. Also shown are the $\sigma_\text{eff}$ value (half the
width of the narrowest interval
containing 68.3\% of the invariant mass distribution) and the full width at
half maximum
(FWHM).}
  \label{fig:results_sigmodels}
\end{figure*}

The final fit function for each category is obtained by summing the functions
for all production
modes normalised to the expected signal yields in that category.
Figure~\ref{fig:results_sigmodels} shows the signal model corresponding to $\mH
= 125\GeV$ for the best resolution category,
 which is the untagged events with the highest signal-to-background ratio and
the highest diphoton BDT score, `Untagged 0'.
Also shown in the same figure is the signal model for the sum of all
categories, with each category weighted by the corresponding
S/(S+B) ratio, where S is the number of signal events, and B is the number of
background events in a window around the \mH peak.
In the figure the effective width ($\sigma_\text{eff}$), defined as half of
the
smallest interval that contains 68.3\% of the invariant mass distribution, is
given, as is the full width at half maximum (FWHM).

\subsection{Background model}
\label{sec:results_background}

The model used to describe the background for each of the analysis categories
is obtained from data using the discrete profiling
method~\cite{DiscreteProfilingMethod}.
In this method, a large set of candidate function families is considered,
including exponential functions, Bernstein polynomials, Laurent series, and
power law functions. These are fit to the \mgg distribution in the mass range
of 100 to 180\GeV.
For each family of functions, a Fisher test~\cite{fisher} is performed to
determine the maximum order to be used in the fit, while the minimum order is
determined by placing a requirement on the goodness of the fit to the data.
The choice of the background function is treated as a discrete nuisance
parameter in the fit to account for the uncertainty associated with the
arbitrary choice of the function.

\section{Systematic uncertainties}
\label{sec:systematics}

The systematic uncertainties are treated differently depending on their effect
on the diphoton invariant mass distributions in the different signal
categories. The systematic uncertainties in the photon energy scale and
resolution modify the shape of the diphoton mass distribution in the signal
model. Other systematic uncertainties, while not affecting the signal shape,
affect the event yield.
The sources of uncertainty included in previous CMS \hgg analyses are
described in Ref.~\cite{HIG-16-040}. A more precise determination of the
systematic uncertainties in the photon energy scale and resolution has been
developed for the present analysis and is described here.

\subsection{Uncertainties in the photon energy scale estimated with electrons}
\label{sec:addsystematics}

The following sources of systematic uncertainties in the photon energy scale
were first estimated using electrons and propagated to the photons.

    \begin{itemize}
        \item \textit{Electron energy scale and resolution}: The uncertainty
in the electron energy scale and resolution corrections are derived using
\Ztoee events by varying the distribution of \Rn, the electron selections used
in the derivation of the corrections, and the transverse energy thresholds on
the electron pairs used in the derivation of the corrections. This uncertainty
is 0.05--0.1\% for electrons in the EB, and 0.1--0.3\% for electrons in the
ECAL endcaps.

        \item \textit{Residual \pt dependence of the energy scale correction}:
Since the corrections for the residual differences between data and simulation
were estimated with \Ztoee events ($\langle\PT\rangle\approx 45\GeV$),
applying them to photons with $\langle\PT\rangle\approx 60\GeV$ introduces an
additional systematic error. The degree of nonclosure of the \pt-dependent electron
energy scale corrections, as described in Section 5.1, is used as the
estimate of this source of uncertainty, and is indicated by the band labelled
as nonlinearity in Fig.~\ref{fig:scale_vs_et}. For electrons having $\pt <
80\GeV$, corresponding to all analysis categories except the Untagged 0
category, this uncertainty is 0.075\%. For electrons having \pt greater than
80\GeV, corresponding to the Untagged 0 category, the uncertainty is 0.15\%.
This uncertainty is applied conservatively on the global energy scale and is
correlated among all photon candidates.
    \end{itemize}

\subsection{Uncertainties due to differences between electrons and photons}

\begin{figure}[htpb]
 \centering
 \includegraphics[width=\cmsFigWidth]{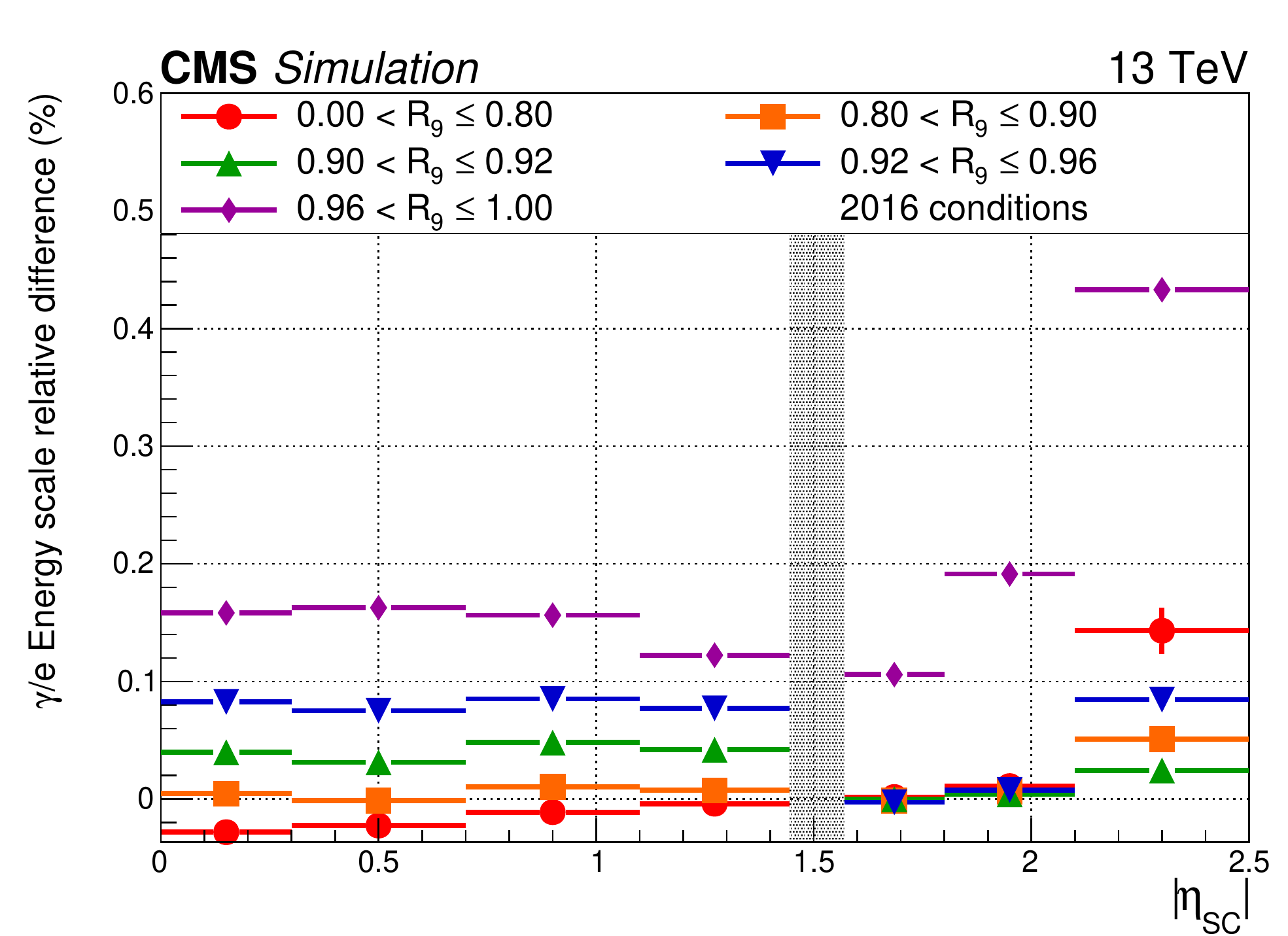}
 \caption{The systematic uncertainty due to the difference between the
electron and photon energy scales from the radiation damage induced
nonuniformity of light collection in ECAL crystals in different supercluster
$\abs{\eta_{SC}}$ and \Rn categories. The method used to evaluate this
uncertainty is described in Section~8.2.
   }
 \label{fig:fnuf_systematics_plots}
\end{figure}

    Additional systematic uncertainties due to the differences between the
response of ECAL to electrons and photons were studied and assigned as
follows:

    \begin{itemize}
        \item \textit{Modelling of the material budget}: The uncertainty in
the material budget between the interaction point and the ECAL, which affects
electron and photon showers differently, was evaluated as described in
Ref.~\cite{HIG-16-040}, and is at most 0.24\% of the photon energy scale.

        \item \textit{Nonuniformity of the light collection}: The shower
maximum for photons is deeper than that of electrons by approximately one
radiation length, which is 0.89\cm in lead tungstate. Hence the
differences in the light collection efficiency along the length of the ECAL
crystals will introduce a difference in the ECAL response to electrons and
photons. To account for this, an additional systematic uncertainty is assigned
to the photon energy scale. Due to the increase in the radiation damage to the
ECAL crystals in Run 2 compared to Run 1, the impact of the nonuniformity in
light collection efficiency has become more important. Therefore, a special
effort has been made to study this effect and to better estimate the
associated systematic uncertainty in the photon energy scale. This is
estimated using a light collection efficiency model derived from a detailed optical
simulation~\cite{SLITRANI} and validated with measurements made with
irradiated crystals~\cite{Adams_2016}. This model takes into account the nonuniformity of the
collection of scintillation light due to radiation damage and the crystal
geometry.
This uncertainty
has been evaluated as a function of \PT, supercluster $\abs{\eta_{SC}}$, and
\Rn using the radiation damage conditions experienced in the 2016 data
taking period. The results are summarised in
Fig.~\ref{fig:fnuf_systematics_plots}. The effect is less than 0.16\% in the
barrel and less than 0.45\% in the endcap, and  affects photons with $\Rn >
0.96$ the most.
The uncertainty is assumed to be correlated among the different $\abs{\eta}$
and \Rn bins but uncorrelated between the barrel and endcap regions due to the
difference in the degree of radiation damage and crystal size.

        \item \textit{Mis-modelling of the input variables to the energy
correction}: The uncertainty in the photon energy scale due to imperfect
modelling of the shower shape in the simulation is found to be negligible
(less than 10\MeV) as a result
 of the good agreement between data and simulation in the different input
variables  used in the photon energy regression correction.
    \end{itemize}

\subsection{Impact of the sources of uncertainty}

    The contribution of each source of the photon energy scale systematic
uncertainty to the total uncertainty in the \mH measurement was evaluated by
performing a likelihood scan removing all but that source and subtracting the
statistical uncertainty in quadrature. The results are summarised in Table
\ref{tab:syst_per_source}. The leading sources of systematic uncertainty
affecting \mH are the residual \PT dependence of the photon energy scale,
nonuniformity of light collection, and the electron energy scale and
resolution correction. The impact of all other sources of systematic
uncertainty were found to be negligible.

\begin{table*}[htbp]
	\centering
        \topcaption{The observed impact of the different
uncertainties on the measurement of \mH}
         \label{tab:syst_per_source}
        \begin{tabular}{l c}
        \hline
            Source & Contribution (\GeVns)  \\ \hline
             Electron energy scale and resolution corrections & 0.10 \\
             Residual \pt dependence of the photon energy scale & 0.11 \\
             Modelling of the material budget & 0.03 \\
             Nonuniformity of the light collection & 0.11 \\
             Total systematic uncertainty & 0.18 \\[\cmsTabSkip]
             Statistical uncertainty & 0.18 \\
             Total uncertainty  & 0.26 \\ \hline
        \end{tabular}
\end{table*}

\section{Results}
\label{sec:results}

\begin{figure*}
    \centering
    \includegraphics[width=0.45\textwidth]{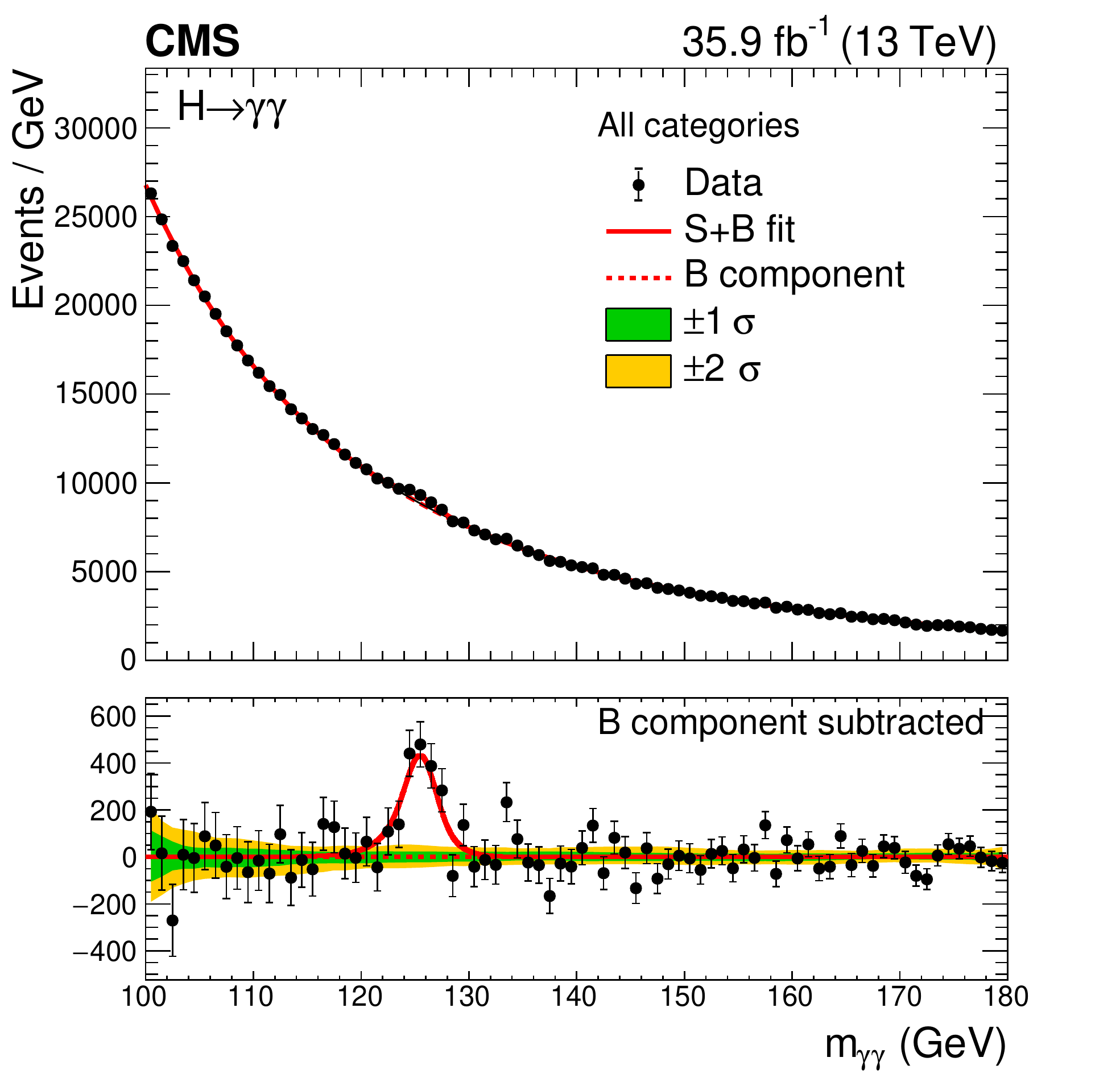}
    \includegraphics[width=0.45\textwidth]{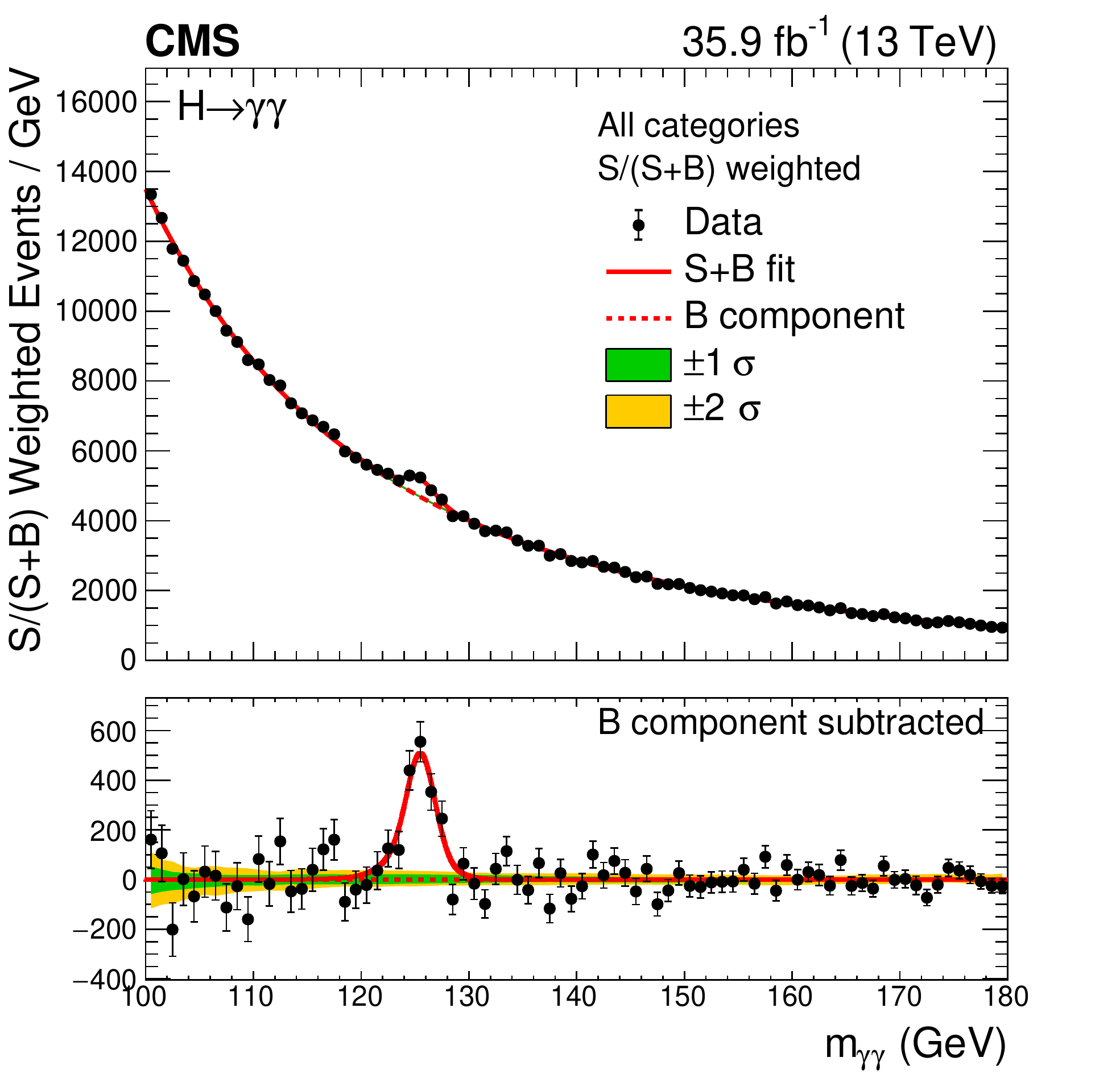}
    \caption{ Data and signal-plus-background model fit for all categories
summed (left) and
where the categories are summed weighted by their corresponding sensitivities,
given by S/(S+B) (right).  The one (green) and
two (yellow) standard deviation bands include the uncertainties in the
background component
of the fit. The lower panel in each plot shows the residuals after the
background subtraction. }
    \label{fig:SplusBFits}
\end{figure*}

To extract the measured value of \mH and its uncertainty, a binned maximum
likelihood fit is performed simultaneously to the
\mgg distributions of the seven analysis categories described in
Sec.~\ref{sec:categories}, in  the  range  $100 < \mgg <
180\GeV$.
We use binned fits to reduce computation time and a bin size of
0.125\GeV, which is small compared to the diphoton mass resolution.
The data and the signal-plus-background model fit for the sum of all analysis
categories is
shown in Fig.~\ref{fig:SplusBFits}. 

\begin{figure*}
 \centering
 \includegraphics[width=0.90\textwidth]{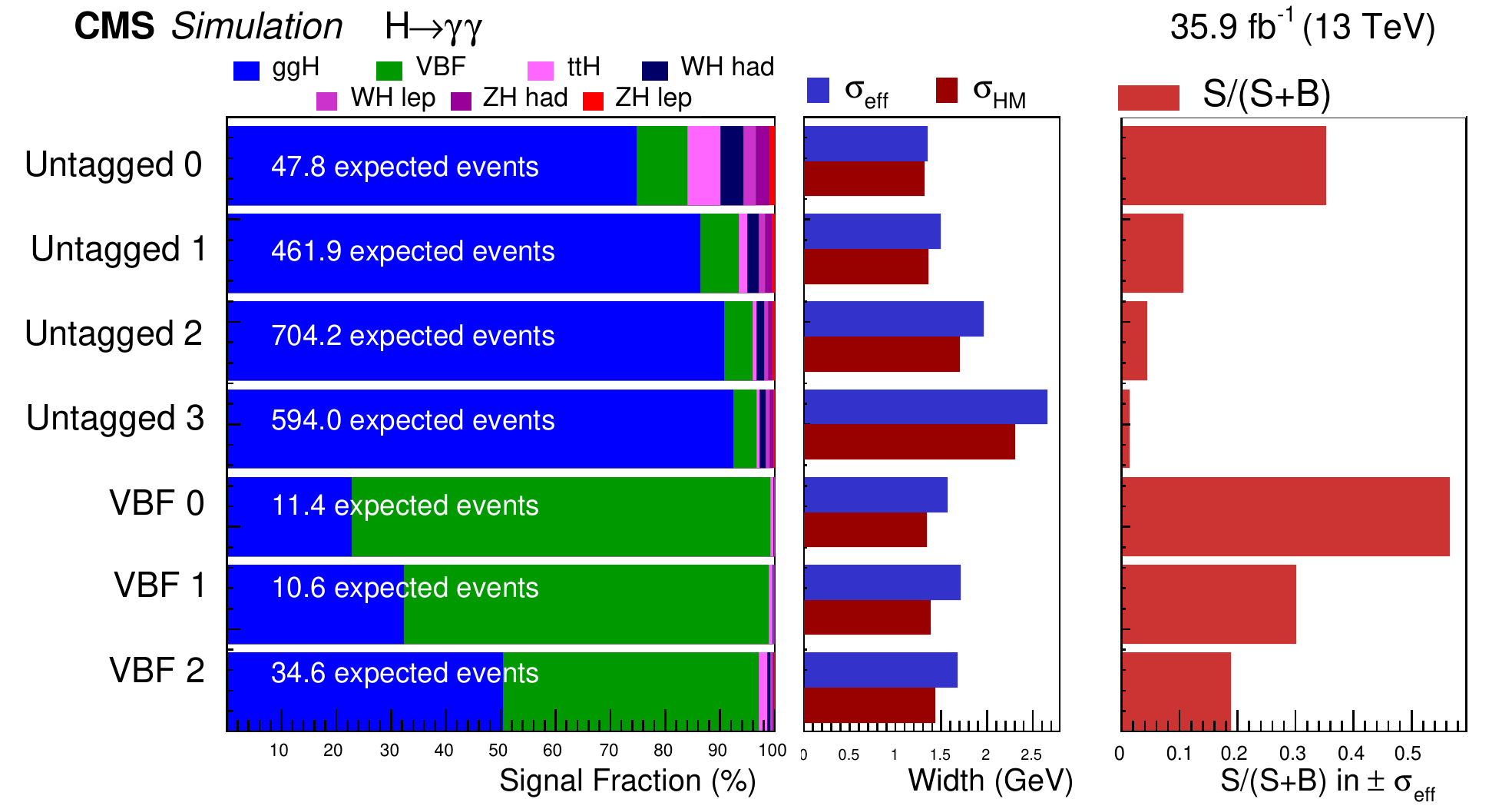}
 \caption{ The expected number of signal events per category and the
percentage
   breakdown per production mode. The $\sigma_\text{eff}$ value (half the
width of the narrowest interval
containing 68.3\% of the invariant mass distribution) is also shown as an
   estimate of the $m_{\gamma\gamma}$ resolution in that category and compared
   directly to the $\sigma_\text{HM}$. The ratio of the number of signal
   events (S) to the number of signal plus background events (S+B) is shown on
   the right-hand panel.
   }
 \label{fig:statAnalysisColourPlot}
\end{figure*}

The expected number of signal events for each category is summarised in
Fig.~\ref{fig:statAnalysisColourPlot},
 where the contribution of each production mode to each analysis category is
shown. The $\sigma_\text{eff}$ and $\sigma_\text{HM}$ are also listed; the
latter is the FWHM, divided by 2.35.

In the likelihood scan of \mH, other parameters of the
signal and background models are allowed to vary.  Systematic uncertainties
are included in the
form of nuisance parameters, and the results are
obtained using an asymptotic approach~\cite{CMS-NOTE-2011-005}
with  a  test  statistic  based  on  the  profile  likelihood
ratio~\cite{TestStatistic}.
In the fit to extract \mH, two independent signal
strengths for the (\ggH, \ttH) $\to \gamma\gamma$ and (VBF, \VH) $\to \gamma\gamma$ processes are free to  vary.
The best-fit mass of \mH is observed to be
$\mH = 125.78 \pm 0.18\stat \pm 0.18\syst\GeV$, while
it was expected to
have a statistical uncertainty of $\pm$0.21\GeV and a systematic
uncertainty of $\pm$0.18\GeV. The signal strengths obtained were found to
be compatible with the same from  previous analysis in the
diphoton decay channel~\cite{HIG-16-040}.
The expected uncertainties in the measurement were obtained by generating an
Asimov data set~\cite{TestStatistic} from the expected signal from the
standard model plus best-fit background model.
The difference between the measured values of \mH in the \hgg channel in the
two LHC run periods, Run 1~\cite{Run1HGG} and 2016,
 is $\Delta\mH = 1.12 \pm 0.43\GeV$. The compatibility of these two results
is at the level of 2.6 standard deviations. A detailed set of cross-checks
was performed to ensure that this shift is statistical.

\subsection{Combination with the \texorpdfstring{\hzz}{H to ZZ to 4l} mass measurement in the 2016 and Run 1 data sets}
\label{sec:Combination_Run2}

The results of this mass measurement were combined with a measurement of the
same quantity in the \hzz decay channel with the 2016 data set
 reported by CMS in Ref.~\cite{Sirunyan:2017exp}
using the same data set with a preliminary set of detector conditions.

In the combination a possible correlation may exist between electron and photon energy scales.
In the \hgg decay channel, the largest contribution to the uncertainty on the
photon energy scale is due to the difference in the calorimeter response to
electrons and photons, which is only applied to the \hgg decay channel.
Other differences between the two decay channels in the derivation of the
energy scale corrections
 are the much finer binning in \Rn and their
\PT-dependence in the \hgg decay channel.
Additionally the average energy of the electrons in the \hzz decay channel is
much lower than the most probable photon energy in the \hgg decay channel.
Thus we treat the uncertainties, residual to the electron-photon difference,
in the electron and photon energy scales to be uncorrelated between the two
channels.

\begin{figure}
    \centering
    \includegraphics[width=\cmsFigWidth]{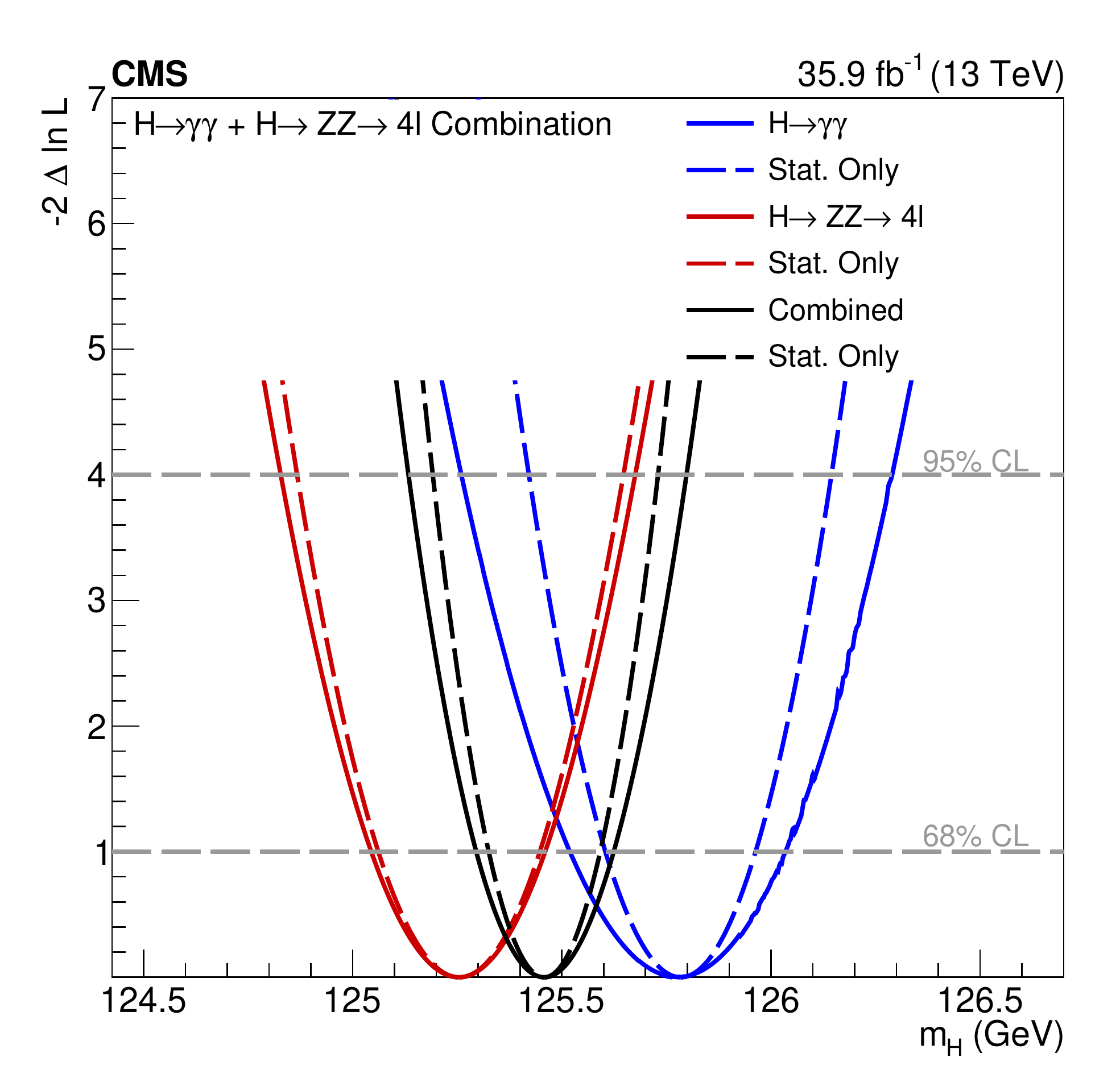}
    \caption{ The  likelihood scan of the measured Higgs boson mass in the
\hgg and \hzz  decay channels individually and for the combination with the
2016 data set. The solid lines are for the full likelihood scan including all
systematic uncertainties, while the dashed lines denote the same with the
statistical uncertainty only.}
    \label{fig:money_mass_scan}
\end{figure}

The combined value of \mH measured from the 2016 data set is observed to
be $\mH = 125.46 \pm 0.13\stat \pm 0.10\syst\GeV$
with an expected
 statistical uncertainty of $\pm$0.16\GeV and an expected systematic uncertainty of
$\pm$0.10\GeV.
Three independent signal strengths for the (\ggH, \ttH) $\to
\gamma\gamma$, (VBF, \VH) $\to \gamma\gamma$ and $\Pp\Pp \to \hzz$
processes are free to vary in the fit to extract \mH, so that we are not
completely dependent on the standard model for the production and decay
ratios.
This result is
in good agreement with the ATLAS+CMS Run 1 measurement~\cite{Aad:2015zhl}, $\mH
= 125.09 \pm 0.24\GeV$.
A scan of the value of twice the negative logarithm of the likelihood
($-2\Delta\ln{L}$) as a function of \mH for the two individual decay channels,
 as well as their combination is shown in Fig.~\ref{fig:money_mass_scan}.

\begin{figure}
    \centering
    \includegraphics[width=\cmsFigWidth]{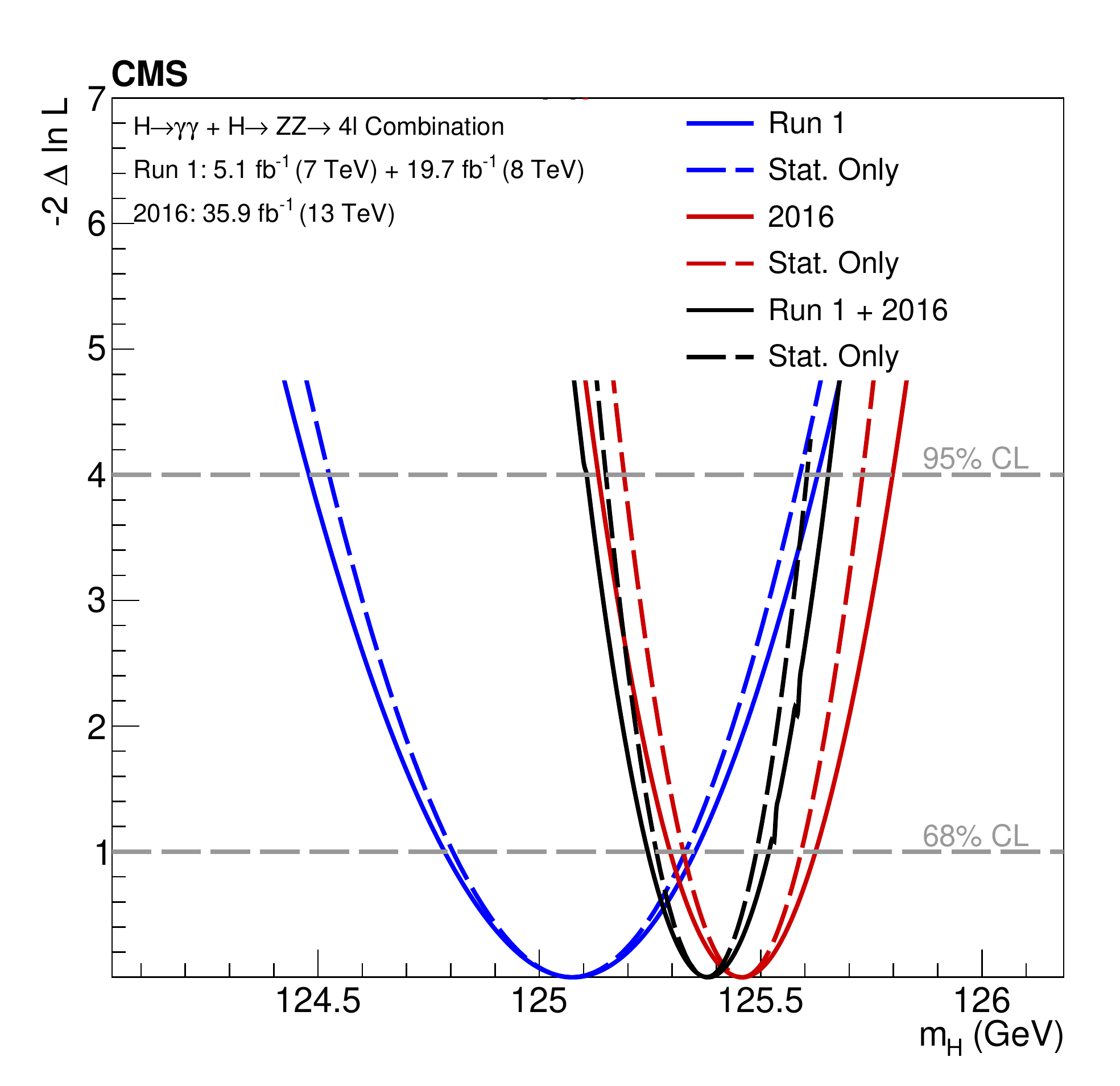}
    \caption{ The  likelihood scan of the combined Higgs boson mass in the
\hgg and \hzz  decay channels with the Run 1 and 2016 data sets and the same
combining the two data sets. The solid lines are for the full likelihood scan
including all systematic uncertainties, while the dashed lines denote the same
with the statistical uncertainty only.}
    \label{fig:money_mass_scan_run1_plus_run2}
\end{figure}

\begin{figure}
 \centering
 \includegraphics[width=\cmsFigWidth]{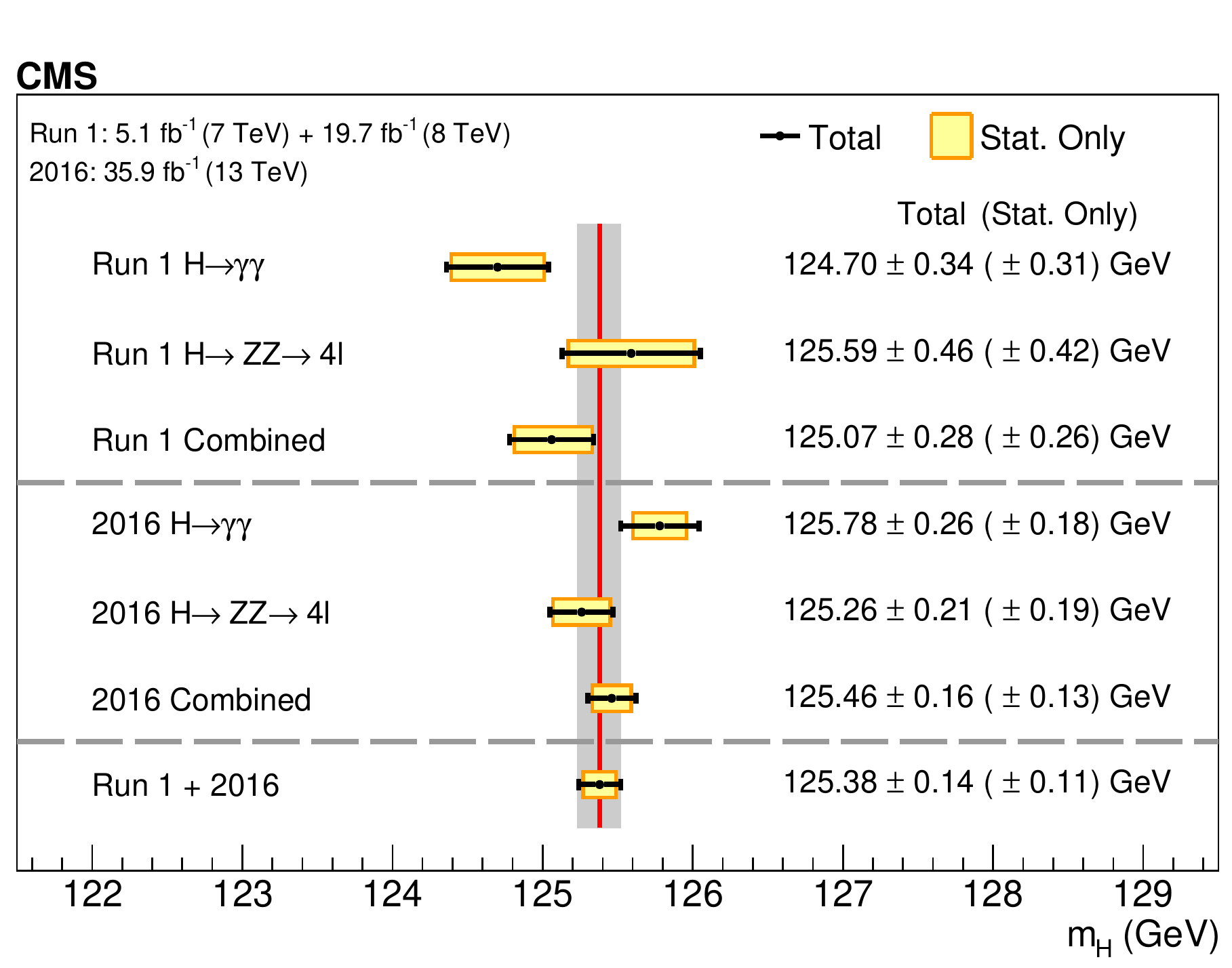}
 \caption{ A summary of the measured Higgs boson mass in the \hgg and \hzz
decay channels, and for the combination of the two is presented here. The
statistical (wider, yellow-shaded bands), and total (black error bars)
uncertainties are indicated.  The (red) vertical line and corresponding (grey)
shaded column indicate the central value and the total uncertainty of the Run
1 + 2016 combined measurement, respectively.
   }
 \label{fig:Mass_Money_Full}
\end{figure}

The same procedure was used to combine this result from the 2016 data set with
the same
 measurement (\hgg and \hzz) obtained from the Run 1 data~\cite{Run1Combo}.
The result of combining the measurements
from both data taking periods is $\mH = 125.38 \pm 0.11\stat \pm
0.08\syst\GeV$ with an expected
 statistical uncertainty of $\pm$0.13\GeV and an expected systematic uncertainty of
$\pm$0.08\GeV. Figure~\ref{fig:money_mass_scan_run1_plus_run2} shows the
likelihood scans of the combined Higgs boson mass in the \hgg and \hzz  decay
channels with the Run 1 and 2016 data sets individually and the same combining
the two data sets. A summary of the individual and combined measurements with
the Run 1 and 2016 data sets is shown in Fig.~\ref{fig:Mass_Money_Full}.

\section{Summary}
\label{sec:summary}

In this Letter we describe a measurement of the Higgs boson mass in the
diphoton decay channel with 35.9\fbinv of data collected in 2016 at 
$\sqrt{s} = 13\TeV$ at the LHC.
New analysis techniques
have been introduced to improve the precision of the measurement and we have
used a refined detector calibration.
The technique that is new with respect to the previous analysis in the
diphoton decay channel~\cite{HIG-16-040} is the introduction
 of residual energy corrections in much finer bins of $\eta$, \PT and the shower shape
variable \Rn of the electrons from \Ztoee decays, in which the electron showers are
reconstructed as photons.
We have also employed a new method to estimate the systematic uncertainty due
to
changes in the transparency of the crystals in the electromagnetic calorimeter
with radiation damage.
The measured value of the Higgs boson mass in the diphoton decay channel is
found to be
$\mH = 125.78 \pm 0.26\GeV$.
This measurement has been combined with a recent measurement by CMS of the
same quantity
in the \hzz decay channel~\cite{Sirunyan:2017exp} to obtain a value of $\mH =
125.46 \pm 0.16\GeV$.
Furthermore, when the Run 2 result with the 2016 data set is combined with the
same measurement performed in Run 1 at 7 and 8\TeV the value of the
Higgs boson mass is found to be $\mH = 125.38 \pm 0.14\GeV$. This is
currently the most precise measurement of the mass of the
Higgs boson.

\begin{acknowledgments}
    We congratulate our colleagues in the CERN accelerator departments for the excellent performance of the LHC and thank the technical and administrative staffs at CERN and at other CMS institutes for their contributions to the success of the CMS effort. In addition, we gratefully acknowledge the computing centres and personnel of the Worldwide LHC Computing Grid for delivering so effectively the computing infrastructure essential to our analyses. Finally, we acknowledge the enduring support for the construction and operation of the LHC and the CMS detector provided by the following funding agencies: BMBWF and FWF (Austria); FNRS and FWO (Belgium); CNPq, CAPES, FAPERJ, FAPERGS, and FAPESP (Brazil); MES (Bulgaria); CERN; CAS, MoST, and NSFC (China); COLCIENCIAS (Colombia); MSES and CSF (Croatia); RPF (Cyprus); SENESCYT (Ecuador); MoER, ERC IUT, PUT and ERDF (Estonia); Academy of Finland, MEC, and HIP (Finland); CEA and CNRS/IN2P3 (France); BMBF, DFG, and HGF (Germany); GSRT (Greece); NKFIA (Hungary); DAE and DST (India); IPM (Iran); SFI (Ireland); INFN (Italy); MSIP and NRF (Republic of Korea); MES (Latvia); LAS (Lithuania); MOE and UM (Malaysia); BUAP, CINVESTAV, CONACYT, LNS, SEP, and UASLP-FAI (Mexico); MOS (Montenegro); MBIE (New Zealand); PAEC (Pakistan); MSHE and NSC (Poland); FCT (Portugal); JINR (Dubna); MON, RosAtom, RAS, RFBR, and NRC KI (Russia); MESTD (Serbia); SEIDI, CPAN, PCTI, and FEDER (Spain); MOSTR (Sri Lanka); Swiss Funding Agencies (Switzerland); MST (Taipei); ThEPCenter, IPST, STAR, and NSTDA (Thailand); TUBITAK and TAEK (Turkey); NASU (Ukraine); STFC (United Kingdom); DOE and NSF (USA). 
 
    \hyphenation{Rachada-pisek} Individuals have received support from the Marie-Curie programme and the European Research Council and Horizon 2020 Grant, contract Nos.\ 675440, 752730, and 765710 (European Union); the Leventis Foundation; the A.P.\ Sloan Foundation; the Alexander von Humboldt Foundation; the Belgian Federal Science Policy Office; the Fonds pour la Formation \`a la Recherche dans l'Industrie et dans l'Agriculture (FRIA-Belgium); the Agentschap voor Innovatie door Wetenschap en Technologie (IWT-Belgium); the F.R.S.-FNRS and FWO (Belgium) under the ``Excellence of Science -- EOS" -- be.h project n.\ 30820817; the Beijing Municipal Science \& Technology Commission, No. Z191100007219010; the Ministry of Education, Youth and Sports (MEYS) of the Czech Republic; the Deutsche Forschungsgemeinschaft (DFG) under Germany’s Excellence Strategy -- EXC 2121 ``Quantum Universe" -- 390833306; the Lend\"ulet (``Momentum") Programme and the J\'anos Bolyai Research Scholarship of the Hungarian Academy of Sciences, the New National Excellence Program \'UNKP, the NKFIA research grants 123842, 123959, 124845, 124850, 125105, 128713, 128786, and 129058 (Hungary); the Council of Science and Industrial Research, India; the HOMING PLUS programme of the Foundation for Polish Science, cofinanced from European Union, Regional Development Fund, the Mobility Plus programme of the Ministry of Science and Higher Education, the National Science Center (Poland), contracts Harmonia 2014/14/M/ST2/00428, Opus 2014/13/B/ST2/02543, 2014/15/B/ST2/03998, and 2015/19/B/ST2/02861, Sonata-bis 2012/07/E/ST2/01406; the National Priorities Research Program by Qatar National Research Fund; the Ministry of Science and Education, grant no. 14.W03.31.0026 (Russia); the Programa Estatal de Fomento de la Investigaci{\'o}n Cient{\'i}fica y T{\'e}cnica de Excelencia Mar\'{\i}a de Maeztu, grant MDM-2015-0509 and the Programa Severo Ochoa del Principado de Asturias; the Thalis and Aristeia programmes cofinanced by EU-ESF and the Greek NSRF; the Rachadapisek Sompot Fund for Postdoctoral Fellowship, Chulalongkorn University and the Chulalongkorn Academic into Its 2nd Century Project Advancement Project (Thailand); the Kavli Foundation; the Nvidia Corporation; the SuperMicro Corporation; the Welch Foundation, contract C-1845; and the Weston Havens Foundation (USA). \end{acknowledgments}
\bibliography{auto_generated}

\providecommand{\href}[2]{#2}\begingroup\raggedright\begin{thebibliography}{10}%
\makeatletter
\providecommand{\hrefCMSnoop }[0]{\@secondoftwo}%
\makeatother
\providecommand{\doi}{\texttt{doi:}\begingroup \urlstyle{tt}\Url}

\bibitem{Aad:2012tfa}
\hrefCMSnoop {}{{ATLAS Collaboration}, ``{Observation of a new particle in the
  search for the Standard Model Higgs boson with the ATLAS detector at the
  LHC}'',} \textit{ Phys. Lett. B} \textbf{ 716} (2012) 1,
  \href{http://dx.doi.org/10.1016/j.physletb.2012.08.020}{\doi{10.1016/j.physletb.2012.08.020}},
  \href{http://www.arXiv.org/abs/1207.7214}{\texttt{arXiv:1207.7214}}.

\bibitem{Chatrchyan:2012ufa}
\hrefCMSnoop {}{{CMS Collaboration}, ``Observation of a new boson at a mass of
  125 {GeV} with the {CMS} experiment at the {LHC}'',} \textit{ Phys. Lett. B}
  \textbf{ 716} (2012) 30,
  \href{http://dx.doi.org/10.1016/j.physletb.2012.08.021}{\doi{10.1016/j.physletb.2012.08.021}},
  \href{http://www.arXiv.org/abs/1207.7235}{\texttt{arXiv:1207.7235}}.

\bibitem{CMSLONG}
\hrefCMSnoop {}{{CMS Collaboration}, ``{Observation of a new boson with mass
  near 125 GeV in $\rm{pp}$ Collisions at $\sqrt{s}$ = 7 and 8 TeV}'',}
  \textit{ JHEP} \textbf{ 06} (2013) 081,
  \href{http://dx.doi.org/10.1007/JHEP06(2013)081}{\doi{10.1007/JHEP06(2013)081}},
\href{http://www.arXiv.org/abs/1303.4571}{\texttt{arXiv:1303.4571}}.
%%CITATION = ARXIV:1303.4571;%%.

\bibitem{Aad:2015zhl}
\hrefCMSnoop {}{ATLAS and {CMS Collaborations}, ``{Combined measurement of the
  Higgs boson mass in $\rm{pp}$ collisions at $\sqrt{s}=7$ and 8 TeV with the
  ATLAS and CMS experiments}'',} \textit{ Phys. Rev. Lett.} \textbf{ 114}
  (2015) 191803,
  \href{http://dx.doi.org/10.1103/PhysRevLett.114.191803}{\doi{10.1103/PhysRevLett.114.191803}},
\href{http://www.arXiv.org/abs/1503.07589}{\texttt{arXiv:1503.07589}}.
%%CITATION = ARXIV:1503.07589;%%.

\bibitem{Sirunyan:2017exp}
\hrefCMSnoop {}{{CMS Collaboration}, ``{Measurements of properties of the Higgs
  boson decaying into the four-lepton final state in $\rm{pp}$ collisions at $
  \sqrt{s}=13 $ TeV}'',} \textit{ JHEP} \textbf{ 11} (2017) 047,
  \href{http://dx.doi.org/10.1007/JHEP11(2017)047}{\doi{10.1007/JHEP11(2017)047}},
\href{http://www.arXiv.org/abs/1706.09936}{\texttt{arXiv:1706.09936}}.
%%CITATION = ARXIV:1706.09936;%%.

\bibitem{Aaboud:2018wps}
\hrefCMSnoop {}{{ATLAS Collaboration}, ``{Measurement of the Higgs boson mass
  in the $H\rightarrow ZZ^* \rightarrow 4\ell$ and $H \rightarrow \gamma\gamma$
  channels with $\sqrt{s}=13$ TeV $\rm{pp}$ collisions using the ATLAS
  detector}'',} \textit{ Phys. Lett. B} \textbf{ 784} (2018) 345,
  \href{http://dx.doi.org/10.1016/j.physletb.2018.07.050}{\doi{10.1016/j.physletb.2018.07.050}},
\href{http://www.arXiv.org/abs/1806.00242}{\texttt{arXiv:1806.00242}}.
%%CITATION = ARXIV:1806.00242;%%.

\bibitem{Trigger}
\hrefCMSnoop {}{{CMS Collaboration}, ``{The CMS trigger system}'',} \textit{
  JINST} \textbf{ 12} (2017) P01020,
  \href{http://dx.doi.org/10.1088/1748-0221/12/01/P01020}{\doi{10.1088/1748-0221/12/01/P01020}},
\href{http://www.arXiv.org/abs/1609.02366}{\texttt{arXiv:1609.02366}}.
%%CITATION = ARXIV:1609.02366;%%.

\bibitem{Chatrchyan:2008zzk}
\hrefCMSnoop {}{{CMS Collaboration}, ``The {CMS} experiment at the {CERN}
  {LHC}'',} \textit{ JINST} \textbf{ 3} (2008) S08004,
\href{http://dx.doi.org/10.1088/1748-0221/3/08/S08004}{\doi{10.1088/1748-0221/3/08/S08004}}.
%%CITATION = JINST,3,S08004;%%.

\bibitem{HIG-16-040}
\hrefCMSnoop {}{{CMS Collaboration}, ``{Measurements of Higgs boson properties
  in the diphoton decay channel in proton-proton collisions at $\sqrt{s} =$ 13
  TeV}'',} \textit{ JHEP} \textbf{ 11} (2018) 185,
  \href{http://dx.doi.org/10.1007/JHEP11(2018)185}{\doi{10.1007/JHEP11(2018)185}},
\href{http://www.arXiv.org/abs/1804.02716}{\texttt{arXiv:1804.02716}}.
%%CITATION = ARXIV:1804.02716;%%.

\bibitem{Run1HGG}
\hrefCMSnoop {}{{CMS Collaboration}, ``{Observation of the diphoton decay of
  the Higgs boson and measurement of its properties}'',} \textit{ Eur. Phys. J.
  C} \textbf{ 74} (2014) 3076,
  \href{http://dx.doi.org/10.1140/epjc/s10052-014-3076-z}{\doi{10.1140/epjc/s10052-014-3076-z}},
\href{http://www.arXiv.org/abs/1407.0558}{\texttt{arXiv:1407.0558}}.
%%CITATION = ARXIV:1407.0558;%%.

\bibitem{Agostinelli:2002hh}
\hrefCMSnoop {}{{GEANT4} Collaboration, ``{\GEANTfour}---a simulation
  toolkit'',} \textit{ Nucl. Instrum. Meth. A} \textbf{ 506} (2003) 250,
\href{http://dx.doi.org/10.1016/S0168-9002(03)01368-8}{\doi{10.1016/S0168-9002(03)01368-8}}.
%%CITATION = NUIMA,A506,250;%%.

\bibitem{AMCAT}
J.~Alwall\hrefCMSnoop {}{ {et~al.}, ``{The automated computation of tree-level
  and next-to-leading order differential cross sections, and their matching to
  parton shower simulations}'',} \textit{ JHEP} \textbf{ 07} (2014) 079,
  \href{http://dx.doi.org/10.1007/JHEP07(2014)079}{\doi{10.1007/JHEP07(2014)079}},
\href{http://www.arXiv.org/abs/1405.0301}{\texttt{arXiv:1405.0301}}.
%%CITATION = ARXIV:1405.0301;%%.

\bibitem{Pythia8}
\hrefCMSnoop {}{T.~Sj{\"{o}}strand, S.~Mrenna, and P.~Skands, ``{A brief
  introduction to PYTHIA 8.1}'',} \textit{ Comput. Phys. Commun.} \textbf{ 178}
  (2007) 852,
  \href{http://dx.doi.org/10.1016/j.cpc.2008.01.036}{\doi{10.1016/j.cpc.2008.01.036}},
  \href{http://www.arXiv.org/abs/0710.3820}{\texttt{arXiv:0710.3820}}.

\bibitem{CUETP8}
\hrefCMSnoop {}{{CMS Collaboration}, ``{Event generator tunes obtained from
  underlying event and multiparton scattering measurements}'',} \textit{ Eur.
  Phys. J. C} \textbf{ 76} (2016) 155,
  \href{http://dx.doi.org/10.1140/epjc/s10052-016-3988-x}{\doi{10.1140/epjc/s10052-016-3988-x}},
\href{http://www.arXiv.org/abs/1512.00815}{\texttt{arXiv:1512.00815}}.
%%CITATION = ARXIV:1512.00815;%%.

\bibitem{SHERPA}
\hrefCMSnoop {}{{Sherpa} Collaboration, ``{Event Generation with Sherpa
  2.2}'',} \textit{ SciPost Phys.} \textbf{ 7} (2019) 034,
  \href{http://dx.doi.org/10.21468/SciPostPhys.7.3.034}{\doi{10.21468/SciPostPhys.7.3.034}},
\href{http://www.arXiv.org/abs/1905.09127}{\texttt{arXiv:1905.09127}}.
%%CITATION = ARXIV:1905.09127;%%.

\bibitem{Mad5}
J.~Alwall\hrefCMSnoop {}{ {et~al.}, ``{MadGraph 5: going beyond}'',} \textit{
  JHEP} \textbf{ 06} (2011) 128,
  \href{http://dx.doi.org/10.1007/JHEP06(2011)128}{\doi{10.1007/JHEP06(2011)128}},
\href{http://www.arXiv.org/abs/1106.0522}{\texttt{arXiv:1106.0522}}.
%%CITATION = ARXIV:1106.0522;%%.

\bibitem{Supercluster}
\hrefCMSnoop {}{{CMS Collaboration}, ``{Energy calibration and resolution of
  the CMS electromagnetic calorimeter in $\rm{pp}$ collisions at $\sqrt{s} = 7$
  TeV}'',} \textit{ JINST} \textbf{ 8} (2013) P09009,
  \href{http://dx.doi.org/10.1088/1748-0221/8/09/P09009}{\doi{10.1088/1748-0221/8/09/P09009}},
\href{http://www.arXiv.org/abs/1306.2016}{\texttt{arXiv:1306.2016}}.
%%CITATION = ARXIV:1306.2016;%%.

\bibitem{Regression}
\hrefCMSnoop {}{{CMS Collaboration}, ``{Performance of photon reconstruction
  and identification with the CMS detector in proton-proton Collisions at
  $\sqrt{s}$ = 8 TeV}'',} \textit{ JINST} \textbf{ 10} (2015) P08010,
  \href{http://dx.doi.org/10.1088/1748-0221/10/08/P08010}{\doi{10.1088/1748-0221/10/08/P08010}},
\href{http://www.arXiv.org/abs/1502.02702}{\texttt{arXiv:1502.02702}}.
%%CITATION = ARXIV:1502.02702;%%.

\bibitem{DiscreteProfilingMethod}
\hrefCMSnoop {}{P.~D. Dauncey, M.~Kenzie, N.~Wardle, and G.~J. Davies,
  ``{Handling uncertainties in background shapes: the discrete profiling
  method}'',} \textit{ JINST} \textbf{ 10} (2015) P04015,
  \href{http://dx.doi.org/10.1088/1748-0221/10/04/P04015}{\doi{10.1088/1748-0221/10/04/P04015}},
  \href{http://www.arXiv.org/abs/1408.6865}{\texttt{arXiv:1408.6865}}.

\bibitem{fisher}
\hrefCMSnoop {}{R.~A. Fisher, ``On the interpretation of $\chi^2$ from
  contingency tables, and the calculation of {P}'',} \textit{ J. Roy. Statis.
  Soc.} \textbf{ 85} (1922) 87,
  \href{http://dx.doi.org/10.2307/2340521}{\doi{10.2307/2340521}}.

\bibitem{SLITRANI}
\hrefCMSnoop {}{{F. Gentit}, ``Litrani: a general purpose {Monte-Carlo} program
  simulating light propagation in isotropic or anisotropic media'',} \textit{
  Nucl. Inst. Meth. {A}} \textbf{ 486} (2002) 35,
  \href{http://dx.doi.org/10.1016/S0168-9002(02)00671-X}{\doi{10.1016/S0168-9002(02)00671-X}}.

\bibitem{Adams_2016}
\hrefCMSnoop {}{T.~Adams {et~al.}, ``Beam test evaluation of electromagnetic
  calorimeter modules made from proton-damaged {PbWO$_4$} crystals'',} \textit{
  JINST} \textbf{ 11} (2016) P04012,
  \href{http://dx.doi.org/10.1088/1748-0221/11/04/p04012}{\doi{10.1088/1748-0221/11/04/p04012}}.

\bibitem{CMS-NOTE-2011-005}
\href {https://cds.cern.ch/record/1379837}{ATLAS and {CMS Collaborations},
  ``Procedure for the {LHC} {Higgs} boson search combination in {Summer}
  2011'',} Technical Report CMS-NOTE-2011-005. ATL-PHYS-PUB-2011-11, 2011.

\bibitem{TestStatistic}
\hrefCMSnoop {}{G.~Cowan, K.~Cranmer, E.~Gross, and O.~Vitells, ``{Asymptotic
  formulae for likelihood-based tests of new physics}'',} \textit{ Eur. Phys.
  J. C} \textbf{ 71} (2011) 1554,
  \href{http://dx.doi.org/10.1140/epjc/s10052-011-1554-0}{\doi{10.1140/epjc/s10052-011-1554-0}},
  \href{http://www.arXiv.org/abs/1007.1727}{\texttt{arXiv:1007.1727}}.
[Erratum: \DOI{10.1140/epjc/s10052-013-2501-z}].
%%CITATION = ARXIV:1007.1727;%%.

\bibitem{Run1Combo}
\hrefCMSnoop {}{{CMS Collaboration}, ``{Precise determination of the mass of
  the Higgs boson and tests of compatibility of its couplings with the standard
  model predictions using proton collisions at 7 and 8 $\,\text {TeV}$}'',}
  \textit{ Eur. Phys. J. C} \textbf{ 75} (2015) 212,
  \href{http://dx.doi.org/10.1140/epjc/s10052-015-3351-7}{\doi{10.1140/epjc/s10052-015-3351-7}},
\href{http://www.arXiv.org/abs/1412.8662}{\texttt{arXiv:1412.8662}}.
%%CITATION = ARXIV:1412.8662;%%.

\end{thebibliography}\endgroup
\cleardoublepage \appendix\section{The CMS Collaboration \label{app:collab}}\begin{sloppypar}\hyphenpenalty=5000\widowpenalty=500\clubpenalty=5000\vskip\cmsinstskip
\textbf{Yerevan Physics Institute, Yerevan, Armenia}\\*[0pt]
A.M.~Sirunyan$^{\textrm{\dag}}$, A.~Tumasyan
\vskip\cmsinstskip
\textbf{Institut f\"{u}r Hochenergiephysik, Wien, Austria}\\*[0pt]
W.~Adam, F.~Ambrogi, T.~Bergauer, M.~Dragicevic, J.~Er\"{o}, A.~Escalante~Del~Valle, M.~Flechl, R.~Fr\"{u}hwirth\cmsAuthorMark{1}, M.~Jeitler\cmsAuthorMark{1}, N.~Krammer, I.~Kr\"{a}tschmer, D.~Liko, T.~Madlener, I.~Mikulec, N.~Rad, J.~Schieck\cmsAuthorMark{1}, R.~Sch\"{o}fbeck, M.~Spanring, W.~Waltenberger, C.-E.~Wulz\cmsAuthorMark{1}, M.~Zarucki
% ...
\vskip\cmsinstskip
\dag: Deceased\\
1:  Also at Vienna University of Technology, Vienna, Austria\\
% ...
\end{sloppypar}
%%% END EDITABLE REGION %%%
\end{document}